\documentclass[twocolumn,preprintnumbers,
showpacs,aps,prl]{revtex4}
\usepackage{amsmath,amssymb,amsthm}
\usepackage{amsfonts,bm,latexsym,dcolumn}
\def\be{\begin{equation}}
\def\ee{\end{equation}}
\def\bea{\begin{eqnarray}}
\def\eea{\end{eqnarray}}
\def\p{\partial}

\begin{document}

\preprint{arXiv:0709.4074v1 [hep-th]}

\title{Comment on ``Hawking Radiation and Covariant Anomalies''}

\author{Shuang-Qing Wu} \email{sqwu@phy.ccnu.edu.cn}
\author{Zhan-Yue Zhao}
\affiliation{\centerline{College of Physical Science and Technology,
Central China Normal University, Wuhan, Hubei 430079, People's Republic of China}}

\begin{abstract}
We comment on the consistence of the epsilon anti-symmetric tensor adopted in [R. Banerjee
and S. Kulkarni, arXiv:0707.2449] when it is generalized in the general case where $\sqrt{-g}
\neq 1$. It is pointed out that the correct non-minimal consistent gauge and gravitational
anomalies should by multiplied a factor $\sqrt{-g} \neq 1$. We also sketch the generalization
of their work to the $\sqrt{-g} \neq 1$ case.
\end{abstract}


\maketitle

\textbf{Comment on ``Hawking Radiation and Covariant Anomalies''}

In a recent paper \cite{BK}, Banerjee and Kulkarni suggested that it is conceptually clean
and economical to use only covariant gauge and gravitational anomalies to derive Hawking
radiation from charged black holes \cite{IUW}. Here we point out that while their work is
self-consistent in the case of the metric determinant $\sqrt{-g} = \sqrt{-g_{tt}g_{rr}} = 1$,
a potential ambiguity will arise in the definition of the epsilon anti-symmetric tensor when
$\sqrt{-g}\neq 1$. We then remove this vagueness and present the correct expressions for the
non-minimal consistent gauge and gravitational anomalies. Subsequently, we sketch the generalization
of their work to derive the charge flux and energy-momentum flow in the most general diagonal
metric case where $\sqrt{-g} \neq 1$.

Without loss of generality, let's consider the most general two-dimensional non-extremal metric
given by
\be
ds^2 = f(r)dt^2 -h(r)^{-1}dr^2 \, .
\label{2dem}
\ee
Here we assume $f(r_+) = h(r_+) = 0$ and $\sqrt{-g} = \sqrt{f/h}$ is regular at the horizon
$r = r_+$. Then we define the two dimensional antisymmetric tensors, $\varepsilon^{\mu\nu} =
\epsilon^{\mu\nu}/\sqrt{-g}$ and $\varepsilon_{\mu\nu} = \sqrt{-g}\epsilon_{\mu\nu}$ for the
upper and the lower cases, respectively, together with $\epsilon^{01} = \epsilon_{10} = 1$.
In terms of the usual Pauli matrices $\sigma_i$'s, the $\gamma$-matrices and $\gamma_5$-matrix
are explicitly given by, $\gamma^0 = \sigma_2/\sqrt{f}$, $\gamma^1 = i\sqrt{h}\sigma_1$, $\gamma_0
= \sqrt{f}\sigma_2$, $\gamma_1 = -i\sigma_1/\sqrt{h}$, and $\gamma_5 \equiv \gamma_0\gamma_1/\sqrt{-g}
= -\sigma_3$.

Using the property of $\gamma$-matrices, $\gamma_{5}\gamma^{\mu} = -\varepsilon^{\mu\nu} \gamma_{\nu}$,
it is easy to check that the chiral current
\be
\mathcal{J}^{\mu} = \bar\psi \Big(\frac{1 \pm \gamma_{5}}{2}\Big)\gamma^{\mu}\psi
= \bar\psi \Big(\frac{g^{\mu\nu} \mp \varepsilon^{\mu\nu}}{2}\Big)\gamma_{\nu}\psi \, ,
\ee
satisfies a constraint condition, $\varepsilon_{\alpha\mu}\mathcal{J}^{\mu} = \mp \mathcal{J}_{\alpha}$.

The minimal form of the consistent gauge anomaly is
\be
\nabla_{\mu}J^{\mu} = \pm \frac{e^2}{4\pi\sqrt{-g}}\epsilon^{\alpha\beta}
\p_{\alpha}A_{\beta} = \pm \frac{e^2}{4\pi} \varepsilon^{\alpha\beta}
\p_{\alpha}A_{\beta} \, ,
\ee
where $+(-)$ corresponds to left(right)-handed fields, respectively. Adding a local polynomial to
the consistent current $J^{\mu}$, then the current $\bar J^{\mu} = J^{\mu} +e^2A^{\mu}/(4\pi)$ is
still consistent and should obey the anomalous equation
\be
\nabla_{\mu}\bar J^{\mu} = \pm \frac{e^2}{4\pi \sqrt{-g}}\p_{\alpha}\big[\sqrt{-g}
\big(\varepsilon^{\alpha\beta} \pm g^{\alpha\beta}\big)A_{\beta}\big] \, .
\ee
This is a non-minimal form for the consistent anomaly including the normal parity terms. Both consistent
anomalies will obey the same equation once if the Lorentz gauge condition $\nabla_\mu A^\mu = 0$ is imposed.
The new consistent current can be further modified, by adding a local polynomial, to define a covariant
current
\be
\tilde{J^{\mu}} = \bar J^{\mu} \mp \frac{e^2}{4\pi}A_{\nu}\big(\varepsilon^{\nu\mu}
\pm g^{\nu\mu}\big) \, ,
\ee
which yields the gauge covariant anomaly
\be
\nabla_{\mu}\tilde{J^{\mu}} = \pm\frac{e^2}{4\pi\sqrt{-g}}\epsilon^{\alpha\beta}F_{\alpha\beta}
= \pm\frac{e^2}{4\pi}\varepsilon^{\alpha\beta}F_{\alpha\beta} \, .
\ee

The minimal form of the consistent gravitational anomaly, for right handed fields, is
\be
\nabla_{\mu}T^{\mu}_{~\nu} = \frac{\epsilon^{\beta\delta}}{96\pi\sqrt{-g}}
\p_{\delta}\p_{\alpha}\Gamma^{\alpha}_{\nu\beta}
= \frac{\varepsilon^{\beta\delta}}{96\pi} \p_{\delta}\p_{\alpha}
\Gamma^{\alpha}_{\nu\beta} \, .
\ee
Including the normal parity terms, the non-minimal form for the consistent gravitational anomaly should
read
\be
\nabla_{\mu}\bar T^{\mu}_{~\nu} = \frac{1}{96\pi\sqrt{-g}}
\p_{\delta}\p_{\alpha}\Big[\sqrt{-g}\big(\varepsilon^{\beta\delta}
+g^{\beta\delta}\big)\Gamma^{\alpha}_{\nu\beta}\Big] \, .
\ee

Special care should be taken for the anti-symmetric tensor $\hat{\epsilon}_{\mu\nu}$ in the covariant
gravitational anomaly
\be
\nabla_{\mu}\tilde{T}^{\mu}_{~\nu} = \frac{1}{96\pi\sqrt{-g}}\hat{\epsilon}_{\mu\nu}\p^{\mu}R \, ,
\ee
where $\hat{\epsilon}_{\mu\nu} \equiv g_{\mu\alpha}g_{\nu\beta}\epsilon^{\alpha\beta} = \sqrt{-g}
\varepsilon_{\mu\nu} = -g\epsilon_{\mu\nu}$ should be understood, in accordance with the conventions
adopted in \cite{IUW,WP}.

Finally, the classical energy momentum tensor is traceless, due to an identity $T^{\lambda}_{~\nu}
= \varepsilon^{\lambda\mu}T_{\mu\nu}$, following from the chirality constraint.

The derivation of Hawking flux via covariant anomalies is completely parallel with the $\sqrt{-g} = 1$
case except some modifications as follows. All $\tilde{J}^r$ and $\tilde{T}_{~t}^r$ appearing in \cite{BK}
should be replaced by $\sqrt{-g} \tilde{J}^r$ and $\sqrt{-g}\tilde{T}_{~t}^r$, respectively. Accordingly,
the operator $\nabla_\mu$ in Eqs. (22) and (31) should be multiplied by the factor $\sqrt{-g}$ also.

The surface gravity calculated by this method is
\bea
\kappa &=& \sqrt{-48\pi\tilde{N}^r_{~t}(r_+)} \, , \\
\tilde{N}^r_{~t} &=& \frac{1}{96\pi}\Big(hf^{\prime\prime}
+\frac{1}{2}h^{\prime}f^{\prime} -\frac{h}{f}f^{\prime 2}\Big) \, .
\eea
Since $f(r_+) = h(r_+) = 0$, we find that $\tilde{N}^r_{~t}(r_+) = \frac{-1}{192\pi}f^{\prime}(r_+)
h^{\prime}(r_+)$, and so $\kappa = \frac{1}{2}\sqrt{f^{\prime}(r_+)h^{\prime}(r_+)}$, in agreement
with that obtained in \cite{WP}.

This work is partially supported by the NSFC under Grant No. 10675051.

\medskip\noindent
Shuang-Qing Wu \email{sqwu@phy.ccnu.edu.cn}
and Zhan-Yue Zhao

\smallskip\noindent
College of Physical Science and Technology, Central China Normal University,
Wuhan, Hubei 430079, People's Republic of China

\smallskip\noindent


\end{document}